\documentclass[%
 reprint,
superscriptaddress,
 amsmath,amssymb,
 aps,
floatfix,
]{revtex4-2}

\usepackage{graphicx}
\usepackage{dcolumn}
\usepackage{bm}
\usepackage{amsmath,amssymb}
\usepackage{textcomp,mathcomp}

\begin{document}

\preprint{APS/123-QED}

\title{Observation of local atomic displacements intrinsic to the double zigzag chain structure of 1$T$-$M$Te$_2$ ($M$ = V, Nb, Ta).}

\author{N. Katayama}\thanks{Corresponding author.}\email{katayama.naoyuki.m5@f.mail.nagoya-u.ac.jp}
\affiliation{Department of Applied Physics, Nagoya University, Aichi 464-8603, Japan}
%
%
\author{Y. Matsuda}
\affiliation{Department of Applied Physics, Nagoya University, Aichi 464-8603, Japan}

\author{K. Kojima}
\affiliation{Department of Applied Physics, Nagoya University, Aichi 464-8603, Japan}

\author{T. Hara}
\affiliation{Department of Applied Physics, Nagoya University, Aichi 464-8603, Japan}

\author{S. Kitou}
\affiliation{RIKEN Center for Emergent Matter Science (CEMS), Wako, Saitama, 351-0198, Japan}
\affiliation{Department of Advanced Materials Science, The University of Tokyo, Kashiwa, 277-8561, Chiba, Japan.}

\author{N. Mitsuishi}
\affiliation{RIKEN Center for Emergent Matter Science (CEMS), Wako, Saitama, 351-0198, Japan}

\author{H. Takahashi}
\affiliation{Division of Materials Physics and Center for Spintronics Research Network (CSRN), Graduate School of Engineering Science, Osaka University, Toyonaka, Osaka 560-8531, Japan}
\affiliation{Spintronics Research Network Division, Institute for Open and Transdisciplinary Research Initiatives, Osaka University, Suita, Osaka, 565-0871, Japan}

\author{S. Ishiwata}
\affiliation{Division of Materials Physics and Center for Spintronics Research Network (CSRN), Graduate School of Engineering Science, Osaka University, Toyonaka, Osaka 560-8531, Japan}
\affiliation{Spintronics Research Network Division, Institute for Open and Transdisciplinary Research Initiatives, Osaka University, Suita, Osaka, 565-0871, Japan}

\author{K. Ishizaka}
\affiliation{RIKEN Center for Emergent Matter Science (CEMS), Wako, Saitama, 351-0198, Japan}
\affiliation{Quantum-Phase Electronics Center and Department of Applied Physics, The University of Tokyo, Bunkyo-ku, Tokyo, 113–8656, Japan}

\author{H. Sawa}					
\affiliation{Department of Applied Physics, Nagoya University, Aichi 464-8603, Japan}
\date{\today}

\begin{abstract}

We describe the existence of local distortion discovered in the synchrotron X-ray single crystal structure analysis of layered ditelluride 1$T$-$M$Te$_2$ ($M$ = V, Nb, Ta). In 1$T$-TaTe$_2$, the double zigzag chain structure of Ta is deformed at about 170 K, and heptamer molecules are formed periodically at low temperatures. We found that some of the Ta atoms that compose the double zigzag chain structure appearing at high temperatures are locally displaced, resulting in local dimerization. This tendency weakens when Ta is replaced by V or Nb. Our results indicate that the local distortion persistently survives in these ditellurides, where the electronic degrees of freedom, including orbitals, are weakened. We further discuss the origin of local distortion in these ditellurides, which is different from many usual material systems where molecular formation occurs at low temperatures.

\end{abstract}

\maketitle


\section{\label{sec:level1}Introduction}
Among transition metal compounds with orbital degrees of freedom, there are many substances whose atoms assemble to form ``molecules" at low temperatures \cite{LiRh2O4, CuIr2S4, LiVO2-kj, LiVO2-kj2, LiVS$_2$-1, AlV2O4, LiMoO2, Li033VS2, CsW2O6, Li2RuO3, RuP1, kobayashi, MgTi2O4}. Examples include dimers in LiRh$_2$O$_4$ \cite{LiRh2O4} and CuIr$_2$S$_4$ \cite{CuIr2S4}, trimers in LiVO$_2$ \cite{LiVO2-kj,LiVO2-kj2} and LiVS$_2$  \cite{LiVS$_2$-1}, and heptamer (trimer/tetramer pair) in AlV$_2$O$_4$ \cite{AlV2O4, AlV2O4-2}. Although these molecules that form spontaneously in crystals have been thought to disappear and regular lattices are realized at high temperatures, recent local structure studies have revealed that local lattice distortions appear in various forms in a precursory manner \cite{AlV2O4-2, Li2RuO3_2, LiRh2O4-2, RuP2, CuIr2S4_2, LiVS$_2$-2, MgTi2O4_3}. For example, in Li$_2$RuO$_3$ \cite{Li2RuO3_2} and LiRh$_2$O$_4$ \cite{LiRh2O4-2}, dimers that appear at low temperatures appear as short-range orders at high temperatures; in CuIr$_2$S$_4$ \cite{CuIr2S4_2}, tetragonal distortions appear locally at high temperatures; in LiVS$_2$, short-range orders of zigzag chains that are unrelated to trimer in the low temperature phase appear and slowly fluctuate at high temperatures \cite{LiVS$_2$-2}. These can be interpreted as local nematic states in which the spontaneous symmetry lowering of the electronic system is strongly coupled to the lattice system \cite{CuIr2S4_2}, and the search for various distortion patterns and the elucidation of their mechanisms are important research themes that go beyond the category of molecular formation systems and have a broad impact on physical properties in general.

Layered transition metal ditellurides $M$Te$_2$ provide a unique playground for such studies. In these material systems, Te-Te covalent bonds derived from large tellurium ions occur, which cause the formal valence of Te to shift from 2- and transfer additional electrons to the transition metal element $M$. Depending on the amount of charge transfer, a variety of molecular formation patterns coupled with charge degrees of freedom appear at low temperatures. For example, in IrTe$_2$, a charge-ordered stripe state between Ir$^{3+}$ and Ir$^{4+}$ occurs at low temperatures, forming an Ir$^{4+}$-Ir$^{4+}$ dimer state \cite{IrTe2-1,IrTe2-2}, and superconductivity appears when this dimer phase is suppressed by Pt doping \cite{IrTe2-3}. In 1$T$-$M$Te$_2$ ($M$ = V, Nb, Ta), transition metal elements form quasi-one-dimensional double zigzag chains, named ``ribbon chain", as shown in Fig.~\ref{fig:fig1}(a) \cite{Bronsema, Brown, Whangbo}. This chain is formed from multiple linear trimers, with each $M$ element offering 2/3 electrons for the formation of one trimer \cite{Whangbo, mitsuishi}. An interesting feature of this system is that only at 1$T$-TaTe$_2$ the modulation of the charge changes at low temperatures below $T_c$ $\sim$ 170 K and the double zigzag chain changes to Ta heptamers, as shown in Fig.~\ref{fig:fig1}(b) \cite{Jansen}. This transformation is different from the situation in conventional matter systems where molecular formation occurs from a regular lattice, where the electron degrees of freedom are expected to be highly degenerate. Do unique local distortions appear in such ditellurides at high temperatures? In addition, clarifying whether the nature of local distortion differs between 1$T$-TaTe$_2$, where heptamerization occurs, and 1$T$-NbTe$_2$ and 1$T$-VTe$_2$, where the double zigzag chain is maintained at low temperatures, will provide important insights into the background physics that generates local distortion.

In this article, we report on the structural analysis of 1$T$-$M$Te$_2$ ($M$ = V, Nb, Ta) single crystals using synchrotron radiation X-rays. The anisotropy of the atomic displacement parameters (ADPs) obtained from the structural analysis suggests that the $M$ atom at the center of the double zigzag chain of these diterulides is locally distorted, in the zigzag chain direction. Structural analysis using the split-site model shows that local distortion toward the heptamerization occurs, and this tendency is strongest at 1$T$-TaTe$_2$. This indicates that local distortions appear universally, as in many molecule-forming systems, even if the average structure of the high-temperature phase is in a state where the degeneracy of the orbital degrees of freedom has been resolved.

\section{\label{sec:level2}Results and discussions}
\subsection{\label{sec:level3}Sample Preparation and experimental details}
Single crystal samples of 1$T$-$M$Te$_2$ ($M$ = V, Nb, Ta) were synthesized by a conventional solid phase reaction method. The mixture of the constituent elements in their amphoteric ratios was vacuum-sealed and sintered at 1000 $\tccentigrade$ for $M$ = Nb and Ta, and at 850 $\tccentigrade$ for $M$ = V for 15 hours. The obtained samples are basically powders, but some of them contain tiny single crystals of several tens of micrometers on a side, which were used for diffraction experiments using synchrotron radiation x-rays. X-ray diffraction (XRD) experiments were conducted using the BL02B1 beamline at SPring-8 at an x-ray energy of 40 keV. The typical dimensions of the 1$T$-$M$Te$_2$ ($M$ = V, Nb, Ta) single crystal used for the XRD experiment were 20 × 20 × 20 $\mu$m$^3$. A He-gas-blowing device was employed to cool the sample to 100 K. A 2D CdTe PILATUS detector was utilized with the diffractometer. The CrysAlisPro program was used to integrate the diffraction profiles. Diffraction intensity averaging and refinement of structural parameters were performed using Jana2006 program \cite{JANA}. Crystal structure was visualized by using VESTA \cite{vesta}. The obtained powder diffraction data were indexed using Conograph \cite{conograph}, and the analysis was performed using Rietan-FP \cite{RIETAN}.

\begin{figure}
\includegraphics[width=86mm]{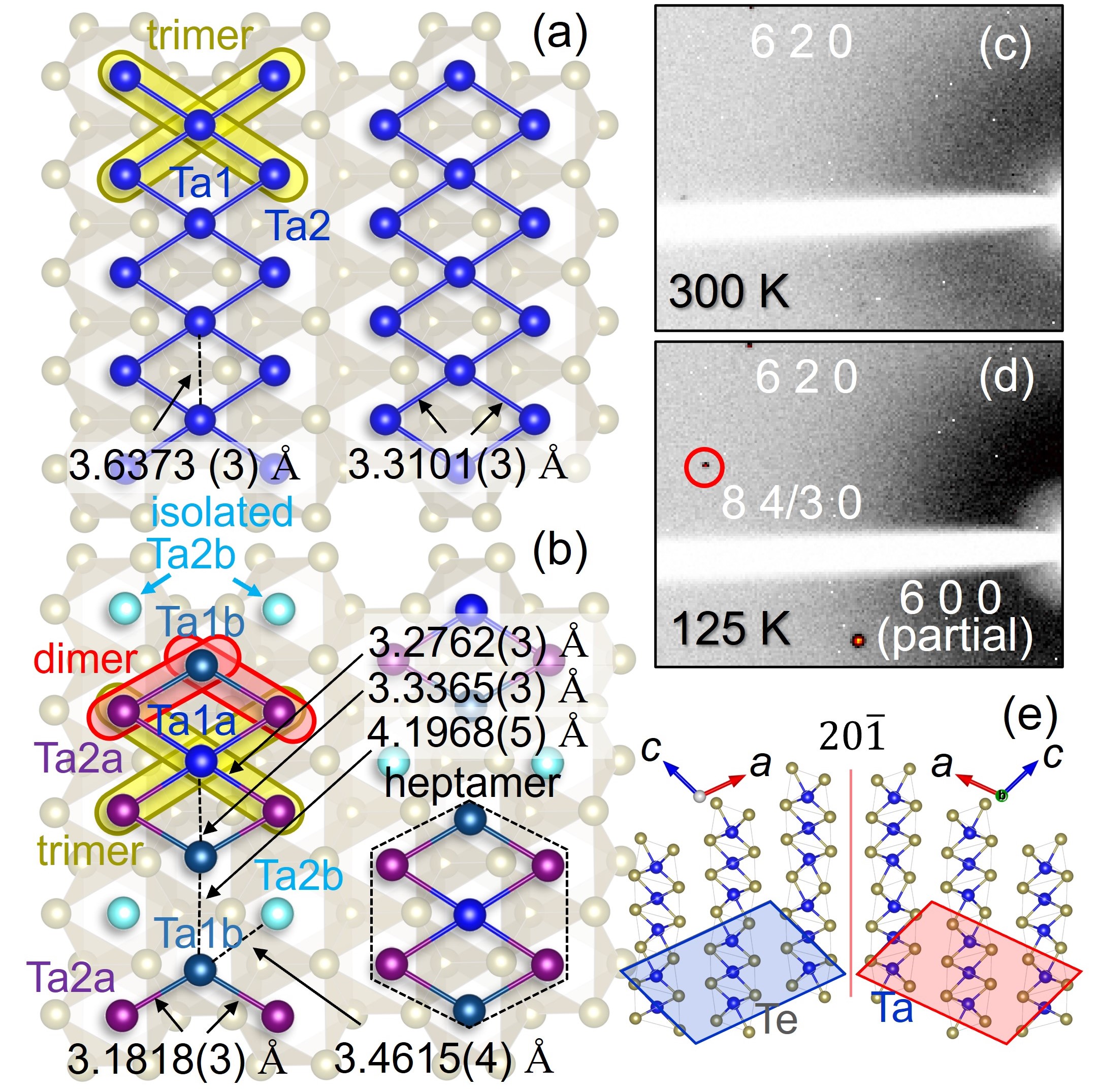}
\caption{\label{fig:fig1} (a-b) The lattice structure of Ta ions at (a) 300 K and (b) 125 K. (c-d) Single crystal XRD patterns of 1$T$-TaTe$_2$ at (c) 300 K and (d) 125 K. Both the high and low temperature phases could be refined using the previously reported structure. Details are shown in the Appendix of this paper. (e) Geometrical arrangement of merohedral domains sharing the 20$\bar{1}$ plane in the crystal structure of the high-temperature phase monoclinic $C$2/$m$.}
\end{figure}


\subsection{\label{sec:level4}Structural studies of 1$T$-TaTe$_2$}

Fig.~\ref{fig:fig1}(c) and (d) show single-crystal XRD patterns of 1$T$-TaTe$_2$ in the high temperature (300 K) and low temperature (125 K) phases. Although the Bragg peaks remain sharp, the presence of a merohedral domain sharing a 20$\bar{1}$ plane, as shown in Fig.~\ref{fig:fig1}(e), is confirmed. In the following, we show the results of the analysis in which only one domain component is extracted. The diffraction pattern changes below the phase transition around 170 K, and the superlattice peak shown by the circle is observed at low temperatures. 

The obtained crystal structures are shown in Figs.~\ref{fig:fig1}(a) and (b), both of which are consistent with the previously reported structures \cite{Brown, Jansen, TaTe2_synchrotron}. Details of the structure analysis results are summarized in the Appendix of this paper. It is noteworthy that the Ta-Ta distance constituting the linear trimer changes significantly between the high and low temperature phases. In the high-temperature phase, the linear trimer is composed of equally spaced Ta2-Ta1-Ta2 arrays, whereas in the low-temperature phase, the displacement of Ta1b ions is accompanied by a large difference in the distance between adjacent Ta-Ta inside the Ta2a-Ta1b-Ta2b array. This indicates that the Ta2a-Ta1b-Ta2b array does not form a linear trimer in the low-temperature phase, but rather transforms into a Ta2a-Ta1b dimer and isolated Ta2b ion. Although the contraction of the Ta-Ta distance upon heptamerization is greatest between Ta1a and Ta1b, the Ta1a-Ta1b distance of $\sim$3.34 \AA~in the low temperature phase is still much longer than the Ta2a-Ta1b distance of $\sim$3.18 \AA, indicating that the Ta2a-Ta1b bond is more essential. The large change in the Ta1a-Ta1b distance associated with the phase transition should be a side effect of the $b$ axis displacement of Ta1b to form dimers with two Ta2a simultaneously.

\begin{figure}
\includegraphics[width=86mm]{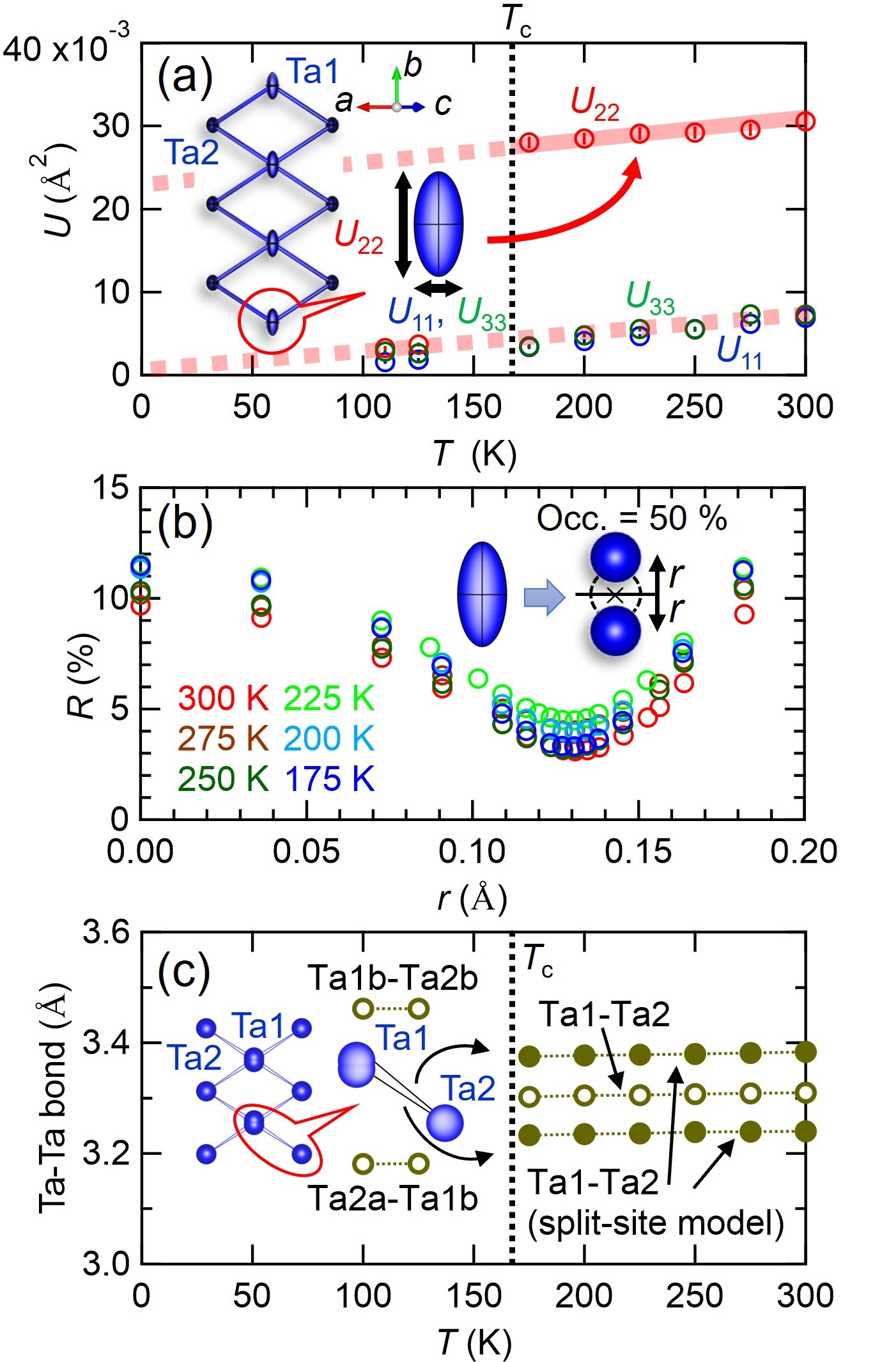}
\caption{\label{fig:fig2} (a) Temperature dependence of the atomic displacement parameters $U_{11}$, $U_{22}$, and $U_{33}$ of the Ta1 site in the high temperature phase. The equations relating the temperature factor T to these atomic displacement parameters are given in the Appendix of this paper. $U_{11}$, $U_{22}$, $U_{33}$ are the mean square amplitudes $\langle$$u^2$$\rangle$ in the reciprocal lattice vector $a^*$, $b^*$, $c^*$ directions. The inset shows the thermal oscillating ellipsoid showing 99 \% probability of the Ta1 site. (b) Temperature dependence of $R$ values at each $r$ obtained by the split-site model. The inset shows a schematic picture of the split-site model. (c) Temperature dependence of Ta-Ta distance for average structure and split-site model, respectively.
}
\end{figure}

Structural analysis revealed that the atomic positions of the high-temperature phase are almost the same as those previously reported \cite{Brown}, and the anisotropic ADP at the Ta1 site also shows an anomalous elongation in the double zigzag chain direction as shown in the inset of Fig.~\ref{fig:fig2}(a), as indicated in a previous report \cite{Jansen}. Fig.~\ref{fig:fig2}(a) shows that the $U_{22}$ parameter has an unusually large value compared to $U_{11}$ and $U_{33}$. It is important to note that the $U_{11}$ and $U_{33}$ parameters are zero when extrapolated toward 0 K, but the $U_{22}$ parameter clearly reaches a finite value. This indicates that the anomalous increase in $U_{22}$ is not simply anisotropic strong thermal oscillation, but rather a local distortion at the Ta1 site. This is consistent with the possibility of dynamic disorder discussed in the previous report \cite{Jansen}.

In order to clarify the local displacement of the ions at the Ta1 site, we performed a structural analysis using a split-site model. This is an analytical method that examines the change in the confidence factor $R$ with the change in distance $r$, assuming that the Ta1 site ions are not at the central position, but two ions with occupancy 0.5 exist at positions +$r$ and -$r$ along the $b$ axis, as shown in the inset of Fig.~\ref{fig:fig2}(b). If the atomic displacement as represented by this model does not actually occur, the $R$ value will show a minimum at $r$ = 0. If it occurs, the $R$ value will be a minimum at a finite $r$. As shown in Fig.~\ref{fig:fig2}(b), the analysis shows that the $R$ value varies significantly with $r$, reaching a minimum at about $r$ = 0.13 \AA. This indicates that there is an intrinsic local atomic displacement in the high-temperature phase of 1$T$-TaTe$_2$, where the local atomic displacement $r$ value that minimizes the $R$ value is almost independent of temperature, and consequently the Ta1-Ta2 bond that splits into two types, as shown in Fig.~\ref{fig:fig2}(c). It seems strange that such local distortion is independent of temperature, but it should be noted that similar behavior has been observed at high temperatures in AlV$_2$O$_4$ \cite{AlV2O4-2} and Li$_2$RuO$_3$ \cite{Li2RuO3_2} where molecular formation occurs at low temperature.


\begin{figure}
\includegraphics[width=86mm]{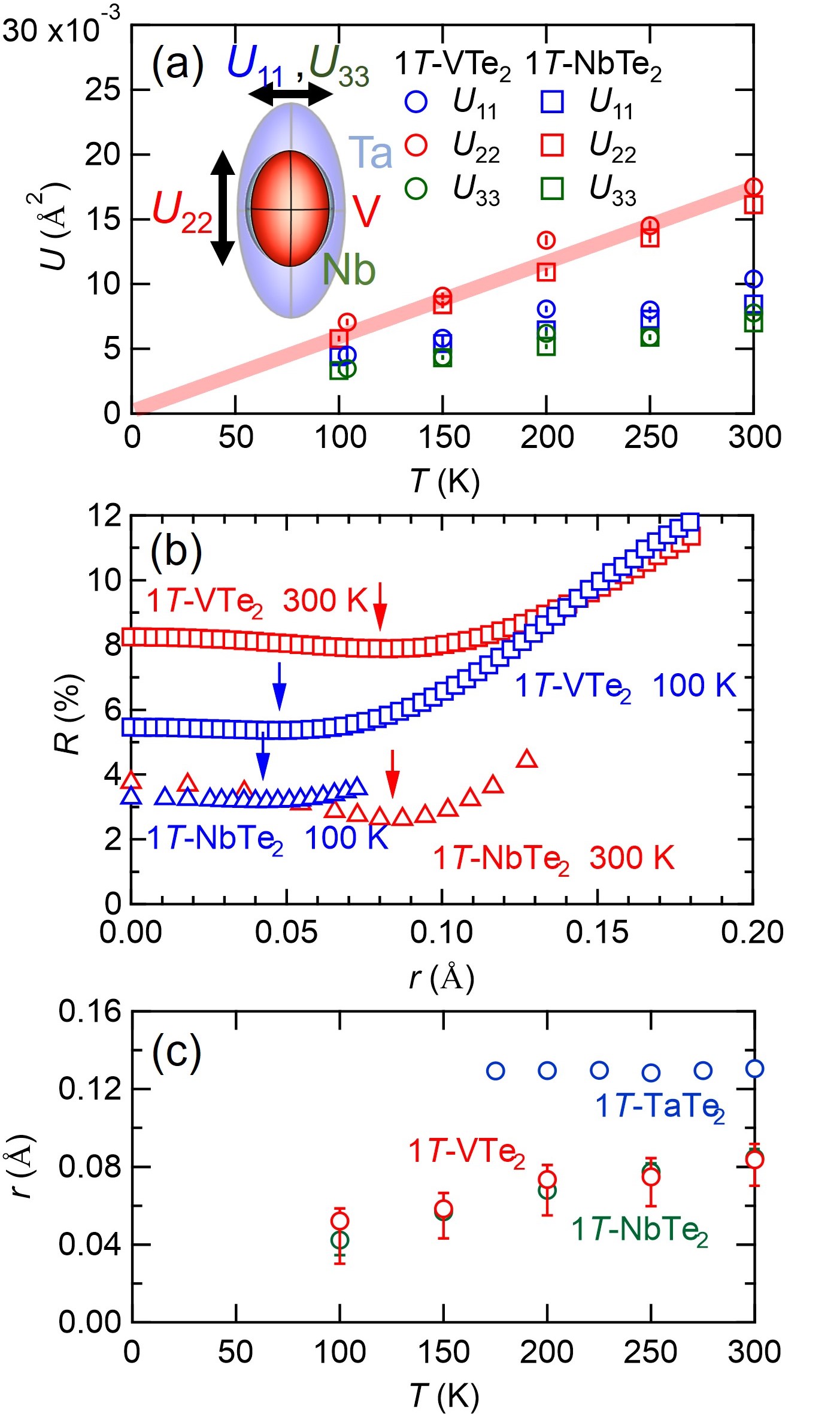}
\caption{\label{fig:fig3} (a) Temperature dependence of the $U_{11}$,$U_{22}$ and $U_{33}$ parameters of 1$T$-NbTe$_2$ and 1$T$-VTe$_2$. Inset shows a schematic picture of the thermal oscillating ellipsoid of 1$T$-NbTe$_2$ and 1$T$-VTe$_2$ compared to that of 1$T$-TaTe$_2$. (b) Temperature dependence of $R$ values at each $r$ obtained by the split-site model. The arrows in the figure indicate the minimum $R$ value. (c) Temperature dependence of local distortion $r$ for 1$T$-NbTe$_2$ and 1$T$-VTe$_2$. The $r$ range from the smallest $R$ value to a 0.5 \% increase was defined as the error bar.}
\end{figure}

\subsection{\label{sec:level4}Structural studies of 1$T$-NbTe$_2$ and 1$T$-VTe$_2$}

Will similar local atomic displacements appear in 1$T$-NbTe$_2$ and 1$T$-VTe$_2$, where the double zigzag chain structure is maintained down to low temperatures? The anisotropic ADPs of the high-temperature phases of these two materials are shown in the inset of Fig.~\ref{fig:fig3}(a). Although a larger $U_{22}$ is realized compared to $U_{11}$ and $U_{33}$, its value is smaller than that of 1$T$-TaTe$_2$. As summarized in Appendix, the values of lattice constants and bond lengths are close between 1$T$-TaTe$_2$, 1$T$-NbTe$_2$ and 1$T$-VTe$_2$. Thus, the difference in magnitude of $U_{22}$ indicates a clear difference in local structure between these telluriums. Also, as shown in Fig.~\ref{fig:fig3}(a), the $U_{22}$ parameter decreases in a temperature-dependent manner to zero at 0 K. Fig.~\ref{fig:fig3}(b) shows the results of the split-site model analysis for these two materials: 1$T$-NbTe$_2$ exhibits a minimum $R$ value at finite $r$, but the value of $r$ that minimizes $R$ at 300 K is about 0.09 \AA, which is smaller than that of 1$T$-TaTe$_2$. The change of $R$ with $r$ is also very small compared to 1$T$-TaTe$_2$. This trend is more pronounced for 1$T$-VTe$_2$, where the $R$ value is almost constant over a wide $r$ range. As shown in Fig.~\ref{fig:fig3}(c), the local atomic displacements, if any, of 1$T$-NbTe$_2$ and 1$T$-VTe$_2$ are much smaller than those of 1$T$-TaTe$_2$ and decrease upon cooling to zero at 0 K.

\subsection{\label{sec:level2}Discussion based on the single-crystal X-ray diffraction experimental results}

\begin{figure}[h]
\includegraphics[width=86mm]{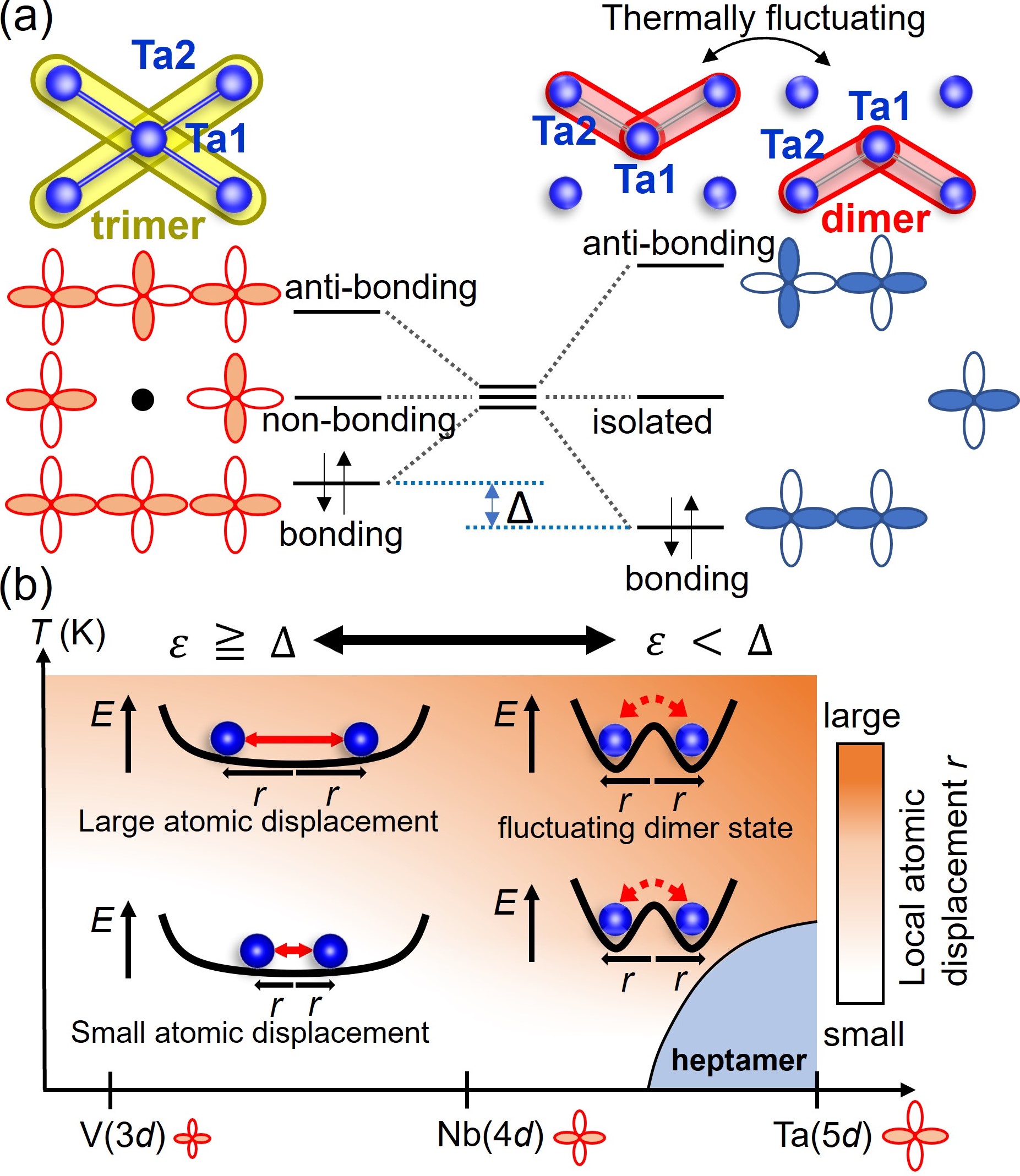}
\caption{\label{fig:fig4} (a) Energy schemes in the linear trimer state and in the dimer and isolated-atom states. (b) Schematic diagram of the elemental $M$ dependence of the relationship between the bonding orbital energy difference $\Delta$ and the lattice system energy loss $\varepsilon$ in the linear trimer state and the dimer with isolated atom states, and the temperature dependence of the local distortion $r$ caused by the difference in $M$.}
\end{figure}

These experimental results raise two questions. The first question is what is the origin of these local atomic displacements. In the high-temperature phase of 1$T$-TaTe$_2$, we can predict that two dimer patterns with equal lattice energy can be realized, as shown in the right side of Fig.~\ref{fig:fig4}(a), which are thermally fluctuating, resulting in a linear trimer in the average structure, as shown in the left side of Fig.~\ref{fig:fig4}(a). In the three-center, two-electron bonding state derived from the linear trimer, three orbitals are formed: a bonding orbital, a non-bonding orbital, and an anti-bonding orbital, with two electrons stored in the bonding orbitals, which are energetically stabilized, as shown in the left energy scheme of Fig.~\ref{fig:fig4}(a). When local atomic displacement occurs, the linear trimer is transformed into a dimer-isolated atom pair, which should result in the formation of a bonding orbital and an anti-bonding orbital, derived from the dimer, and an isolated atomic orbital, as shown in the right energy scheme of Fig.~\ref{fig:fig4}(a). The three energetically equally spaced orbitals formed by the linear trimer and dimer-isolated atom pairs are similar at first glance. However, if the actual interatomic distances of the dimer are sufficiently closer than those of the linear trimer, the bonding orbitals should be $\Delta$ lower in energy for the dimer. If the energy difference $\Delta$ is larger than the energy loss due to lattice distortion, $\varepsilon$, the local dimer is expected to stabilize. Since the energies between the two dimer patterns are equal, we expect the two dimer patterns to appear thermally fluctuating, as shown in Fig.~\ref{fig:fig4}(a). The transition process between the two dimer patterns is via a trimer state, and the low energy of the bonding orbital in this trimer state should facilitate the transition between the two patterns. The long-range ordering of this local distortion at low temperatures leads to the formation of heptamer clusters, as shown in Fig.~\ref{fig:fig1}(b). 

Another question is why different behaviors appear for 1$T$-TaTe$_2$, 1$T$-NbTe$_2$, and 1$T$-VTe$_2$. As shown in Fig.~\ref{fig:fig3}(c), local atomic displacements are largest for Ta(5$d$) and small for Nb(4$d$) and V(3$d$). In 1$T$-TaTe$_2$ with large 5$d$ orbitals, the electronic energy gain $\Delta$ is expected to be the largest because the orbitals overlap more during dimerization than in 1$T$-NbTe$_2$ and 1$T$-VTe$_2$. If $\varepsilon$ $<$ $\Delta$ is realized at 1$T$-TaTe$_2$, the overall energy of the system including lattice energy with respect to atomic positions will be as shown in Fig.~\ref{fig:fig4}(b). Since atomic displacement and the associated local dimerization stabilize the system, and this trend does not change even when the temperature is lowered, the atomic displacement $r$ does not show any temperature dependence, and the local distortion is expected to survive until just above the phase transition. On the other hand, for 4$d$ Nb and 3$d$ V, the orbital overlap is small and the energy is not very stable due to local dimerization. Therefore, $\varepsilon$ $\geq$ $\Delta$ is realized and the energy of the whole system including lattice energy has a flat shape with respect to $r$. The energy drops to a low central position when the temperature is lowered. Because of these differences in energy schemes, it is thought that among the three tellurides, only 1$T$-TaTe$_2$ produces large local distortions, which are maintained at low temperatures and develop into heptamers. From the above discussion, it seems likely that in 1$T$-NbTe$_2$ and 1$T$-VTe$_2$, when the distance between adjacent transition metals is closer, the orbital overlap becomes larger and dimer (heptamer) states such as 1$T$-TaTe$_2$ are formed. It should be noted that it has been reported that applying pressure to 1$T$-NbTe$_2$ induces a structural evolution from the trimeric to the dimeric structure \cite{NbTe2_press}, confirming the validity of this argument. 

An argument similar to the local distortion of 1$T$-TaTe$_2$ might be applied to other material systems that exhibit molecular formation at low temperatures, such as the spinel compound AlV$_2$O$_4$. It has been argued that vanadium in AlV$_2$O$_4$ spontaneously forms a heptamer molecule at about 700 K. Based on the Curie paramagnetic component in the low-temperature phase by magnetization measurements, the heptamer has been revealed to be composed of 9 bonds by 18 electrons \cite{AlV2O4}. However, recent structural analysis of the split-site model on synchrotron XRD data has revealed that the heptamer is actually composed of trimer/tetramer pairs \cite{AlV2O4-2}. The two structures are compared in Fig. 25 of the paper by D.I. Khomskii and S.V. Streltsov \cite{KhomskiiSergey}. The nature of the difference between the two types of molecular structures proposed for AlV$_2$O$_4$ can be interpreted as whether the three bonds connecting the upper and lower triangular trimer are linear trimers consisting of three-center two-electron bonds or pairs of dimers and isolated atoms. This is very similar to the present case. An interesting difference is that such local distortions appear in the low-temperature ordered phase in AlV$_2$O$_4$, whereas they appear only in the high-temperature phase in 1$T$-TaTe$_2$. This indicates that unconventional orbital ordering states that appear in the low-temperature phase can universally appear as local distortions in the high-temperature phase, even in systems like 1$T$-TaTe$_2$, where there is no short-range order in the low-temperature ordered phase.

The presence of local distortion in the high-temperature phase not only has important implications for the mechanism of phase transitions, but is also important for understanding the thermodynamics of the high-temperature phase. For example, while high thermal conductivity is generally realized in the metallic state due to itinerant electrons, it has been observed that in CuIr$_2$S$_4$, which undergoes insulating transition with molecular formation at low temperatures, the thermal conductivity of the high temperature metallic phase is lower than that of the insulating low temperature phase. It is argued that local distortion suppresses phonon thermal conduction \cite{CuIr2S4_thermal1, CuIr2S4_thermal2, CuIr2S4_thermal3}. In addition, the multiple degrees of freedom of electrons are expected to play an important role in the entropy changes associated with molecular formation \cite{Li033VS2, LiVS$_2$-1}. The presence of local distortion will be a new factor, not previously considered, to the concept of electron degrees of freedom in the high-temperature phase. In this respect, the finding of differences in local structure among the systematics of $M$ = V, Nb, and Ta in this study should correspond to providing an attractive stage for discussing the role of local distortion on physical properties through comparison.

\begin{acknowledgments}
The work leading to these results has received funding from the Grant in Aid for Scientific Research (Nos.~JP20H02604, JP21K18599, JP21J21236) and Research Foundation for the Electrotechnology of Chubu. This work was carried out under the Visiting Researcher’s Program of the Institute for Solid State Physics, the University of Tokyo, and the Collaborative Research Projects of Laboratory for Materials and Structures, Institute of Innovative Research, Tokyo Institute of Technology. Single-crystal and powder XRD experiments were conducted at the BL02B1 and BL02B2 of SPring-8, Hyogo, Japan (Proposals No. 2019B1085, 2021A0070, 2021B1261, 2021B1136, 2021B1261, 2022A0304, 2022B0607, 2022B1570, 2022B1582, and 2022B1862), and at the BL5S2 of Aichi Synchrotron Radiation Center, Aichi Science and Technology Foundation, Aichi, Japan (Proposals No. 202202037, 202201033, 202105170, 202104111, 2021L3002, 2021L2002, 2021L1002 and 2020L4002).
\end{acknowledgments}

\appendix
\section{Single crystal X-ray diffraction analysis results}


Here, the temperature factor $T$ is expressed as a function of the atomic displacement parameters $U_{11}$, $U_{22}$, $U_{33}$, $U_{12}$, $U_{13}$ and $U_{23}$, using the following equation,
\begin{equation*}
 \begin{split}
T& = \exp{}\{-2\pi^2(h^2a^{*2}U_{11}+k^2b^{*2}U_{22}+l^2c^{*2}U_{33}\\
& \quad +2hka^*b^*U_{12}+2hla^*c^*U_{13}+2klb^*c^*U_{23})\}.
 \end{split}
\end{equation*}

$U_{11}$, $U_{22}$ and $U_{33}$ are the mean square amplitudes $\langle$$u^2$$\rangle$ in the reciplocal lattice vector $a^{*}$, $b^{*}$ and $c^{*}$ directions.

\begin{table*}[htb]
\centering
\caption{Summary of crystallographic data of 1$T$-TaTe$_2$. }
  \begin{tabular}{|c|c|c|c|}  \hline
Temperature (K) & 110 (low $T$ phase) & \multicolumn{2}{|c|}{175 (high $T$ phase)}  \\ \hline
Wavelength (\AA) & 0.31007 & \multicolumn{2}{|c|}{0.31007} \\ \hline
Crystal dimension ($\mu$m$^3$) & 20$\times$20$\times$20 & \multicolumn{2}{|c|}{20$\times$20$\times$20} \\ \hline
space group & $C$2/$m$ & \multicolumn{2}{|c|}{$C$2/$m$} \\ \hline
$a$ (\AA) & 14.7669(3) & \multicolumn{2}{|c|}{14.7408(4)} \\ \hline
$b$ (\AA) & 10.8702(2) & \multicolumn{2}{|c|}{3.62940(10)} \\ \hline
$c$ (\AA) & 9.2926(2) & \multicolumn{2}{|c|}{9.3287(2)} \\ \hline
$\beta$ ($^{\circ}$) & 110.630(8) & \multicolumn{2}{|c|}{110.864(8)} \\ \hline
$V$ (\AA$^3$) & 1395.99(5) & \multicolumn{2}{|c|}{466.36(3)} \\ \hline
$Z$ & 18 & \multicolumn{2}{|c|}{6} \\ \hline
$F$(000) & 3186 & \multicolumn{2}{|c|}{1062} \\ \hline
(sin$\theta$/$\lambda$)$_{Max}$ (\AA$^{-1}$) & 1.3513 & \multicolumn{2}{|c|}{1.3512} \\ \hline
$N_{Total, obs}$ & 92313 & \multicolumn{2}{|c|}{32489} \\ \hline
$N_{Unique, obs}$ & 13372 & \multicolumn{2}{|c|}{4637} \\ \hline
Average redundancy & 6.9 & \multicolumn{2}{|c|}{7.0} \\ \hline
Completeness & 0.904 & \multicolumn{2}{|c|}{0.894} \\ \hline \hline
\multicolumn{4}{|c|}{Structural analysis using anisotropic displacement parameters} \\ \hline \hline
$R_1$ [\# of reflections] & 0.066 [12475] & \multicolumn{2}{|c|}{0.0467 [4259]} \\ \hline
$R_1$ ($I$ $>$ 4$\sigma$) [\# of reflections] & 0.0445 [9365] & \multicolumn{2}{|c|}{0.0358 [3586]} \\ \hline
GOF [\# of reflections] & 1.035 [12475] & \multicolumn{2}{|c|}{1.057 [4259]} \\ \hline \hline
\multicolumn{4}{|c|}{Structural analysis without using anisotropic displacement parameters$^*$} \\ \hline \hline
split-site model & Not used & Use & Not used \\ \hline
$R_1$ [\# of reflections] & 0.0673 [12475] & 0.0468 [4259] & 0.1318 [4259] \\ \hline
$R_1$ ($I$ $>$ 4$\sigma$) [\# of reflections] & 0.0456 [9365] & 0.0359 [3586] & 0.1152 [3586] \\ \hline
GOF [\# of reflections] & 1.037 [12475] & 1.048 [4259] & 1.093 [4259] \\ \hline
\multicolumn{4}{c}{} \\
  \end{tabular}
 \\
\leftline{* In this analysis, isotropic ADP is used only for the Ta1 site at high temperatures and its derived sites at low temperatures,} 
\leftline{and anisotropic ADP is used for the remaining sites. In this table, the Z value indicates how many molecules of that}
\leftline{molecular composition unit are contained in the crystal lattice, and $F$(000) is the total number of electrons in the unit cell.}
\leftline{As noted in the text, there are merohedral twinning in the present telluride that shares certain planes, and some}
\leftline{peaks are composed of contributions from both domains and cannot be separated. Such peaks were excluded from the} 
\leftline{analysis. The resulting completeness, which is the ratio of the number of experimentally observed Bragg peaks to}
\leftline{the number of Bragg peaks that can be observed in principle, is about 0.9; the same occurs for Nb and V derivatives.}
\leftline{ }

\caption{\label{tab:table1}%
Structural parameters of 1$T$-TaTe$_2$ at 175 K without the split-site model. 
}
\begin{ruledtabular}
\begin{tabular}{ccccccc}
&\multicolumn{5}{c}{atomic coordinates}\\
site & Wyck. & Occ. & $x/a$ & $y/b$ & $z/c$ & $U_{eq}$ (\AA$^{2}$) \\ 
\hline
Ta1&2$d$&1&1/2&-1/2&0&0.01177(4)\\
Ta2&4$i$&1&0.63944(2)&0&0.29083(2)& 0.00437(2)\\
Te1&4$i$&1&0.79655(2)&1/2&0.37790(2)&0.00419(3)\\
Te2&4$i$&1&0.49515(2)&-1/2&-0.30921(2)&0.00431(3)\\
Te3&4$i$&1&0.64883(2)&0&0.01094(2)&0.00551(3)\\
\end{tabular}
\end{ruledtabular}
\leftline{ }
\leftline{ }

\caption{\label{tab:table2}%
Anisotropic atomic displacement parameters of 1$T$-TaTe$_2$ at 175 K without the split-site model. 
}
\begin{ruledtabular}
\begin{tabular}{ccccccc}
site~~ & $U_{11}$ (\AA$^{2}$) & $U_{22}$ (\AA$^{2}$) & $U_{33}$ (\AA$^{2}$) & $U_{12}$ (\AA$^{2}$) & $U_{13}$ (\AA$^{2}$) & $U_{23}$ (\AA$^{2}$) \\ \hline
Ta1&0.00341(6)&0.02829(13)&0.00349(7)&0&0.00109(5)&0\\
Ta2&0.00357(3)&0.00539(4)&0.00348(4)&0&0.00044(3)&0\\
Te1&0.00394(5)&0.00370(5)&0.00420(6)&0&0.00054(4)&0\\
Te2&0.00440(5)&0.00427(5)&0.00416(6)&0&0.00139(5)&0\\
Te3&0.00711(6)&0.00444(5)&0.00429(6)&0&0.00119(5)&0\\
\end{tabular}
\end{ruledtabular}
\end{table*}

\begin{table*}[htb]
\centering
\caption{\label{tab:table1}%
Structural parameters of 1$T$-TaTe$_2$ at 175 K with the split-site model. 
}
\begin{ruledtabular}
\begin{tabular}{ccccccc}
&\multicolumn{5}{c}{atomic coordinates}\\
site & Wyck. & Occ. & $x/a$ & $y/b$ & $z/c$ & $U_{eq}$($U_{iso}$) (\AA$^{2}$) \\ 
\hline
Ta1&4$g$&0.5&1/2&-0.53555(7)&0&0.00430(3)\\
Ta2&4$i$&1&0.63944(2)&0&0.29083(2)& 0.00441(2)\\
Te1&4$i$&1&0.79655(2)&1/2&0.37788(2)&0.00425(3)\\
Te2&4$i$&1&0.49516(2)&-1/2&-0.30919(2)&0.00432(2)\\
Te3&4$i$&1&0.64882(2)&0&0.01091(2)&0.00551(3)\\
\end{tabular}
\end{ruledtabular}
\leftline{ }
\leftline{ In the split-site model analysis, isotropic ADP is used only for the Ta1 site, and anisotropic ADP is used for the other } 
\leftline{elements. The fact that $U_{iso}$ at the Ta1 site and $U_{eq}$ at other sites are almost equal indicates the validity of this analysis.} 
\leftline{}
\leftline{ }

\caption{\label{tab:table2}%
Structural parameters of 1$T$-TaTe$_2$ at 110 K without the split-site model. 
}
\begin{ruledtabular}
\begin{tabular}{ccccccc}
&\multicolumn{5}{c}{atomic coordinates}\\
site & Wyck. & Occ. & $x/a$ & $y/b$ & $z/c$ & $U_{eq}$ (\AA$^{2}$) \\ 
\hline
Ta1a&2$d$&1&0&1/2&1/2&0.00234(3)\\
Ta1b&4$h$&1&0&0.19316(2)&1/2&0.00205(2)\\
Ta2a&8$j$&1&0.85918(2)&0.33810(2)&0.21004(2)&0.00196(2)\\
Ta2b&4$i$&1&0.86247(2)&0&0.20874(2)& 0.00229(2)\\
Te1a&8$j$&1&0.70510(2)&-0.16609(2)&0.12266(2)&0.00228(2)\\
Te1b&4$i$&1&0.79909(2)&0&-0.12305(3)&0.00223(3)\\
Te2a&4$i$&1&0.99543(2)&1/2&0.18923(3)&0.00216(3)\\
Te2b&8$j$&1&0.99419(2)&0.17096(2)&0.19070(2)&0.00226(2)\\
Te3a&4$i$&1&0.86348(2)&0&0.49175(3)&0.00207(3)\\
Te3b&8$j$&1&0.84603(2)&0.33834(2)&0.49008(2)&0.00215(2)\\
\end{tabular}
\end{ruledtabular}
\leftline{ }
\leftline{ }

\caption{\label{tab:table2}%
Anisotropic atomic displacement parameters of 1$T$-TaTe$_2$ at 110 K without the split-site model. 
}
\begin{ruledtabular}
\begin{tabular}{ccccccc}
site~~ & $U_{11}$ (\AA$^{2}$) & $U_{22}$ (\AA$^{2}$) & $U_{33}$ (\AA$^{2}$) & $U_{12}$ (\AA$^{2}$) & $U_{13}$ (\AA$^{2}$) & $U_{23}$ (\AA$^{2}$) \\ \hline
Ta1a&0.00143(5)&0.00343(6)&0.00215(7)&0&0.00064(5)&0\\
Ta1b&0.00172(4)&0.00223(4)&0.00215(5)&0&0.00063(4)&0\\
Ta2a&0.00158(3)&0.00201(3)&0.00199(3)&-0.00004(2)&0.00024(2)&-0.00005(2)\\
Ta2b&0.00187(4)&0.00266(4)&0.00210(5)&0&0.00040(4)&0\\
Te1a&0.00187(4)&0.00207(4)&0.00247(5)&0.00004(3)&0.00023(4)&0.00006(3)\\
Te1b&0.00169(6)&0.00216(6)&0.00242(7)&0&0.00021(5)&0\\
Te2a&0.00208(6)&0.00191(5)&0.00251(7)&0&0.00083(5)&0\\
Te2b&0.00226(4)&0.00199(4)&0.00247(5)&-0.00008(3)&0.00077(4)&0.00001(3)\\
Te3a&0.00197(6)&0.00182(5)&0.00246(7)&0&0.00084(5)&0\\
Te3b&0.00186(4)&0.00195(4)&0.00258(5)&0.00001(3)&0.00072(4)&0.00001(3)\\
\end{tabular}
\end{ruledtabular}
\leftline{ }
\leftline{ Ta1a and Ta1b sites are from the high temperature phase Ta1 site, Ta2a and Ta2b sites are from the high temperature}
\leftline{phase Ta2 site, Te1a and Te1b sites are from the high temperature phase Te1 site, Te2a and Te2b are from the high}
\leftline{temperature phase Te2 site, Te3a and Te3b are from the high temperature phase Te3 site of the high-temperature phase. }
\end{table*}
\leftline{ }

\begin{table*}[htb]
\centering
\caption{Summary of crystallographic data of 1$T$-NbTe$_2$ at 300 K. }
  \begin{tabular}{|c|c|c|}  \hline
Temperature (K) & \multicolumn{2}{|c|}{300}  \\ \hline
Wavelength (\AA) & \multicolumn{2}{|c|}{0.31011} \\ \hline
Crystal dimension ($\mu$m$^3$) & \multicolumn{2}{|c|}{20$\times$20$\times$10} \\ \hline
space group & \multicolumn{2}{|c|}{$C$2/$m$} \\ \hline
$a$ (\AA) & \multicolumn{2}{|c|}{14.6619(5)} \\ \hline
$b$ (\AA) & \multicolumn{2}{|c|}{3.63760(10)} \\ \hline
$c$ (\AA) & \multicolumn{2}{|c|}{9.3144(3)} \\ \hline
$\beta$ ($^{\circ}$) & \multicolumn{2}{|c|}{110.070(8)} \\ \hline
$V$ (\AA$^3$) & \multicolumn{2}{|c|}{466.61(3)} \\ \hline
$Z$ & \multicolumn{2}{|c|}{6} \\ \hline
$F$(000) & \multicolumn{2}{|c|}{870} \\ \hline
(sin$\theta$/$\lambda$)$_{Max}$ (\AA$^{-1}$) & \multicolumn{2}{|c|}{1.2499} \\ \hline
$N_{Total, obs}$ & \multicolumn{2}{|c|}{29479} \\ \hline
$N_{Unique, obs}$ & \multicolumn{2}{|c|}{3618} \\ \hline
Average redundancy & \multicolumn{2}{|c|}{8.1} \\ \hline
Completeness & \multicolumn{2}{|c|}{0.878} \\ \hline \hline
\multicolumn{3}{|c|}{Structural analysis using anisotropic displacement parameters} \\ \hline \hline
$R_1$ [\# of reflections] & \multicolumn{2}{|c|}{0.0434 [3494]} \\ \hline
$R_1$ ($I$ $>$ 4$\sigma$) [\# of reflections] & \multicolumn{2}{|c|}{0.0290 [2778]} \\ \hline
GOF [\# of reflections] & \multicolumn{2}{|c|}{0.981 [3494]} \\ \hline \hline
\multicolumn{3}{|c|}{Structural analysis without using anisotropic displacement parameters$^*$} \\ \hline \hline
split-site model & Use & Not used \\ \hline
$R_1$ [\# of reflections] & $~~~$ 0.0441 [3494] $~~~$ & 0.0564 [3494] \\ \hline
$R_1$ ($I$ $>$ 4$\sigma$) [\# of reflections] &  0.0294 [2778]  & 0.0407 [2778] \\ \hline
GOF [\# of reflections] &  0.998 [3494]  & 1.061 [3494] \\ \hline
  \end{tabular}
\leftline{* In this analysis, isotropic ADP is used only for the Nb1 site, and anisotropic ADP is used for the remaining sites.}

\caption{\label{tab:table1}%
Summary of crystallographic data of 1$T$-NbTe$_2$ at 300 K without the split-site model. 
}
\begin{ruledtabular}
\begin{tabular}{ccccccc}
&\multicolumn{5}{c}{atomic coordinates}\\
site & Wyck. & Occ. & $x/a$ & $y/b$ & $z/c$ & $U_{eq}$ (\AA$^{2}$) \\ 
\hline
Nb1&2$d$&1&1/2&-1/2&0&0.01070(4)\\
Nb2&4$i$&1&0.638910(14)&0&0.29045(2)& 0.00829(3)\\
Te1&4$i$&1&0.797228(11)&1/2&0.378315(18)&0.00888(3)\\
Te2&4$i$&1&0.503383(11)&-1/2&0.309539(18)&0.00867(3)\\
Te3&4$i$&1&0.649411(11)&0&0.009461(18)&0.00865(3)\\
\end{tabular}
\end{ruledtabular}

\caption{\label{tab:table2}%
Anisotropic atomic displacement parameters of 1$T$-NbTe$_2$ at 300 K without the split-site model. 
}
\begin{ruledtabular}
\begin{tabular}{ccccccc}
site~~ & $U_{11}$ (\AA$^{2}$) & $U_{22}$ (\AA$^{2}$) & $U_{33}$ (\AA$^{2}$) & $U_{12}$ (\AA$^{2}$) & $U_{13}$ (\AA$^{2}$) & $U_{23}$ (\AA$^{2}$) \\ \hline
Nb1&0.00864(8)&0.01617(12)&0.00719(9)&0&0.00259(7)&0\\
Nb2&0.00853(6)&0.00850(7)&0.00692(6)&0&0.00147(5)&0\\
Te1&0.00891(5)&0.00769(5)&0.00863(5)&0&0.00119(4)&0\\
Te2&0.00998(5)&0.00792(5)&0.00824(5)&0&0.00327(4)&0\\
Te3&0.00971(5)&0.00735(5)&0.00864(5)&0&0.00284(4)&0\\
\end{tabular}
\end{ruledtabular}

\caption{\label{tab:table1}%
Structural parameters of 1$T$-NbTe$_2$ at 300 K with the split-site model. 
}
\begin{ruledtabular}
\begin{tabular}{ccccccc}
&\multicolumn{5}{c}{atomic coordinates}\\
site & Wyck. & Occ. & $x/a$ & $y/b$ & $z/c$ & $U_{eq}$($U_{iso}$) (\AA$^{2}$) \\ 
\hline
Nb1&4$g$&0.5&1/2&-0.52321(19)&0&0.00812(5)\\
Nb2&4$i$&1&0.63892(2)&0&0.29044(2)& 0.00830(3)\\
Te1&4$i$&1&0.79723(2)&1/2&0.37831(2)&0.00889(3)\\
Te2&4$i$&1&0.50338(2)&-1/2&0.30954(2)&0.00868(3)\\
Te3&4$i$&1&0.64941(2)&0&0.00948(2)&0.00865(3)\\
\end{tabular}
\end{ruledtabular}
\leftline{ }
\leftline{ In the split-site model analysis, isotropic ADP is used only for the Nb1 site, and anisotropic ADP is used for the other } 
\leftline{elements. The fact that $U_{iso}$ at the Nb1 site and $U_{eq}$ at other sites are almost equal indicates the validity of this analysis.} 
\end{table*}

\begin{table*}[htb]
\centering
\caption{Summary of crystallographic data of 1$T$-NbTe$_2$ at 100 K. }
  \begin{tabular}{|c|c|c|}  \hline
Temperature (K) & \multicolumn{2}{|c|}{100}  \\ \hline
Wavelength (\AA) & \multicolumn{2}{|c|}{0.31011} \\ \hline
Crystal dimension ($\mu$m$^3$) & \multicolumn{2}{|c|}{20$\times$20$\times$10} \\ \hline
space group & \multicolumn{2}{|c|}{$C$2/$m$} \\ \hline
$a$ (\AA) & \multicolumn{2}{|c|}{14.5770(3)} \\ \hline
$b$ (\AA) & \multicolumn{2}{|c|}{3.63410(10)} \\ \hline
$c$ (\AA) & \multicolumn{2}{|c|}{9.2961(2)} \\ \hline
$\beta$ ($^{\circ}$) & \multicolumn{2}{|c|}{109.956(8)} \\ \hline
$V$ (\AA$^3$) & \multicolumn{2}{|c|}{462.88(3)} \\ \hline
$Z$ & \multicolumn{2}{|c|}{6} \\ \hline
$F$(000) & \multicolumn{2}{|c|}{870} \\ \hline
(sin$\theta$/$\lambda$)$_{Max}$ (\AA$^{-1}$) & \multicolumn{2}{|c|}{1.2500} \\ \hline
$N_{Total, obs}$ & \multicolumn{2}{|c|}{28124} \\ \hline
$N_{Unique, obs}$ & \multicolumn{2}{|c|}{3608} \\ \hline
Average redundancy & \multicolumn{2}{|c|}{7.8} \\ \hline
Completeness & \multicolumn{2}{|c|}{0.882} \\ \hline \hline
\multicolumn{3}{|c|}{Structural analysis using anisotropic displacement parameters} \\ \hline \hline
$R_1$ [\# of reflections] & \multicolumn{2}{|c|}{0.0453 [3504]} \\ \hline
$R_1$ ($I$ $>$ 4$\sigma$) [\# of reflections] & \multicolumn{2}{|c|}{0.0345 [2842]} \\ \hline
GOF [\# of reflections] & \multicolumn{2}{|c|}{0.986 [3504]} \\ \hline \hline
\multicolumn{3}{|c|}{Structural analysis without using anisotropic displacement parameters$^*$} \\ \hline \hline
split-site model & Use & Not used \\ \hline
$R_1$ [\# of reflections] & $~~~$ 0.0459 [3504] $~~~$ & 0.0469 [3504] \\ \hline
$R_1$ ($I$ $>$ 4$\sigma$) [\# of reflections] &  0.0349 [2842]  & 0.0359 [2842] \\ \hline
GOF [\# of reflections] &  0.993 [3504]  & 1.017 [3504] \\ \hline
  \end{tabular}
\leftline{* In this analysis, isotropic ADP is used only for the Nb1 site, and anisotropic ADP is used for the remaining sites.}

\caption{\label{tab:table1}%
Summary of crystallographic data of 1$T$-NbTe$_2$ at 100 K without the split-site model. 
}
\begin{ruledtabular}
\begin{tabular}{ccccccc}
&\multicolumn{5}{c}{atomic coordinates}\\
site & Wyck. & Occ. & $x/a$ & $y/b$ & $z/c$ & $U_{eq}$ (\AA$^{2}$) \\ 
\hline
Nb1&2$d$&1&1/2&-1/2&0&0.00464(4)\\
Nb2&4$i$&1&0.63853(2)&0&0.20955(2)& 0.00385(3)\\
Te1&4$i$&1&0.79750(2)&1/2&0.37881(2)&0.00392(3)\\
Te2&4$i$&1&0.503383(11)&-1/2&0.30954(2)&0.00391(3)\\
Te3&4$i$&1&0.64991(2)&0&0.00941(2)&0.00385(3)\\
\end{tabular}
\end{ruledtabular}

\caption{\label{tab:table2}%
Anisotropic atomic displacement parameters of 1$T$-NbTe$_2$ at 100 K without the split-site model. 
}
\begin{ruledtabular}
\begin{tabular}{ccccccc}
site~~ & $U_{11}$ (\AA$^{2}$) & $U_{22}$ (\AA$^{2}$) & $U_{33}$ (\AA$^{2}$) & $U_{12}$ (\AA$^{2}$) & $U_{13}$ (\AA$^{2}$) & $U_{23}$ (\AA$^{2}$) \\ \hline
Nb1&0.00440(9)&0.00579(10)&0.00332(9)&0&0.00079(8)&0\\
Nb2&0.00426(7)&0.00336(7)&0.00319(7)&0&0.00030(5)&0\\
Te1&0.00431(6)&0.00301(5)&0.00359(6)&0&0.00025(4)&0\\
Te2&0.00465(6)&0.00315(5)&0.00350(5)&0&0.00086(4)&0\\
Te3&0.00457(6)&0.00290(5)&0.00353(5)&0&0.00069(4)&0\\
\end{tabular}
\end{ruledtabular}

\caption{\label{tab:table1}%
Structural parameters of 1$T$-NbTe$_2$ at 100 K with the split-site model. 
}
\begin{ruledtabular}
\begin{tabular}{ccccccc}
&\multicolumn{5}{c}{atomic coordinates}\\
site & Wyck. & Occ. & $x/a$ & $y/b$ & $z/c$ & $U_{eq}$($U_{iso}$) (\AA$^{2}$) \\ 
\hline
Nb1&4$g$&0.5&1/2&0.5113(4)&0&0.00405(6)\\
Nb2&4$i$&1&0.63854(2)&0&0.29028(2)& 0.00384(4)\\
Te1&4$i$&1&0.79751(2)&1/2&0.37881(2)&0.00391(3)\\
Te2&4$i$&1&0.50338(2)&-1/2&0.30954(2)&0.00390(3)\\
Te3&4$i$&1&0.64991(2)&0&0.00943(2)&0.00383(3)\\
\end{tabular}
\end{ruledtabular}
\leftline{ }
\leftline{ In the split-site model analysis, isotropic ADP is used only for the Nb1 site, and anisotropic ADP is used for the other } 
\leftline{elements. The fact that $U_{iso}$ at the Nb1 site and $U_{eq}$ at other sites are almost equal indicates the validity of this analysis.} 
\end{table*}

\begin{table*}[htb]
\centering
\caption{Summary of crystallographic data of 1$T$-VTe$_2$ at 100 K. }
  \begin{tabular}{|c|c|c|}  \hline
Temperature (K) & \multicolumn{2}{|c|}{100}  \\ \hline
Wavelength (\AA) & \multicolumn{2}{|c|}{0.311} \\ \hline
Crystal dimension ($\mu$m$^3$) & \multicolumn{2}{|c|}{30$\times$30$\times$30} \\ \hline
space group & \multicolumn{2}{|c|}{$C$2/$m$} \\ \hline
$a$ (\AA) & \multicolumn{2}{|c|}{14.3110(2)} \\ \hline
$b$ (\AA) & \multicolumn{2}{|c|}{3.59625(4)} \\ \hline
$c$ (\AA) & \multicolumn{2}{|c|}{9.09130(10)} \\ \hline
$\beta$ ($^{\circ}$) & \multicolumn{2}{|c|}{109.602(2)} \\ \hline
$V$ (\AA$^3$) & \multicolumn{2}{|c|}{440.770(9)} \\ \hline
$Z$ & \multicolumn{2}{|c|}{6} \\ \hline
$F$(000) & \multicolumn{2}{|c|}{762} \\ \hline
(sin$\theta$/$\lambda$)$_{Max}$ (\AA$^{-1}$) & \multicolumn{2}{|c|}{1.7824} \\ \hline
$N_{Total, obs}$ & \multicolumn{2}{|c|}{41843} \\ \hline
$N_{Unique, obs}$ & \multicolumn{2}{|c|}{9196} \\ \hline
Average redundancy & \multicolumn{2}{|c|}{4.6} \\ \hline
Completeness & \multicolumn{2}{|c|}{0.831} \\ \hline \hline
\multicolumn{3}{|c|}{Structural analysis using anisotropic displacement parameters} \\ \hline \hline
$R_1$ [\# of reflections] & \multicolumn{2}{|c|}{0.0708 [8327]} \\ \hline
$R_1$ ($I$ $>$ 4$\sigma$) [\# of reflections] & \multicolumn{2}{|c|}{0.0607 [7414]} \\ \hline
GOF [\# of reflections] & \multicolumn{2}{|c|}{1.323 [8327]} \\ \hline \hline
\multicolumn{3}{|c|}{Structural analysis without using anisotropic displacement parameters$^*$} \\ \hline \hline
split-site model & Use & Not used \\ \hline
$R_1$ [\# of reflections] & $~~~$ 0.0711 [8327] $~~~$ & 0.0720 [8327] \\ \hline
$R_1$ ($I$ $>$ 4$\sigma$) [\# of reflections] & 0.0610 [7414] & 0.0620 [7414] \\ \hline
GOF [\# of reflections] & 1.319 [8327] & 1.316 [8327] \\ \hline
  \end{tabular}
\leftline{* In this analysis, isotropic ADP is used only for the V1 site, and anisotropic ADP is used for the remaining sites.}

\caption{\label{tab:table1}%
Summary of crystallographic data of 1$T$-VTe$_2$ at 100 K without the split-site model. 
}
\begin{ruledtabular}
\begin{tabular}{ccccccc}
&\multicolumn{5}{c}{atomic coordinates}\\
site & Wyck. & Occ. & $x/a$ & $y/b$ & $z/c$ & $U_{eq}$ (\AA$^{2}$) \\ 
\hline
V1&2$d$&1&1/2&-1/2&0&0.00503(6)\\
V2&4$i$&1&0.64217(4)&0&0.29701(6)& 0.00431(4)\\
Te1&4$i$&1&0.79577(2)&1/2&0.37769(2)&0.00396(2)\\
Te2&4$i$&1&0.50775(2)&-1/2&0.30650(2)&0.00377(2)\\
Te3&4$i$&1&0.64568(2)&0&0.01414(2)&0.00378(2)\\
\end{tabular}
\end{ruledtabular}

\caption{\label{tab:table2}%
Anisotropic atomic displacement parameters of 1$T$-VTe$_2$ at 100 K without the split-site model. 
}
\begin{ruledtabular}
\begin{tabular}{ccccccc}
site~~ & $U_{11}$ (\AA$^{2}$) & $U_{22}$ (\AA$^{2}$) & $U_{33}$ (\AA$^{2}$) & $U_{12}$ (\AA$^{2}$) & $U_{13}$ (\AA$^{2}$) & $U_{23}$ (\AA$^{2}$) \\ \hline
V1&0.00440(13)&0.00684(16)&0.00345(13)&0&0.00079(11)&0\\
V2&0.00447(8)&0.00465(9)&0.00319(9)&0&0.00047(7)&0\\
Te1&0.00407(3)&0.00373(3)&0.00334(3)&0&0.00026(3)&0\\
Te2&0.00420(3)&0.00365(3)&0.00297(3)&0&0.00055(3)&0\\
Te3&0.00403(3)&0.00353(3)&0.00327(3)&0&0.00053(3)&0\\
\end{tabular}
\end{ruledtabular}

\caption{\label{tab:table1}%
Structural parameters of 1$T$-VTe$_2$ at 100 K with the split-site model. 
}
\begin{ruledtabular}
\begin{tabular}{ccccccc}
&\multicolumn{5}{c}{atomic coordinates}\\
site & Wyck. & Occ. & $x/a$ & $y/b$ & $z/c$ & $U_{eq}$($U_{iso}$) (\AA$^{2}$) \\ 
\hline
V1&4$g$&0.5&1/2&0.5135(5)&0&0.00428(7)\\
V2&4$i$&1&0.64217(4)&0&0.29696(6)& 0.00431(4)\\
Te1&4$i$&1&0.79577(2)&1/2&0.37769(2)&0.00396(2)\\
Te2&4$i$&1&0.50775(2)&-1/2&0.30650(2)&0.00377(2)\\
Te3&4$i$&1&0.64568(2)&0&0.01414(2)&0.00378(2)\\
\end{tabular}
\end{ruledtabular}
\leftline{ }
\leftline{ In the split-site model analysis, isotropic ADP is used only for the V1 site, and anisotropic ADP is used for the other } 
\leftline{elements. The fact that $U_{iso}$ at the V1 site and $U_{eq}$ at other sites are almost equal indicates the validity of this analysis.} 
\end{table*}

\nocite{*}

\clearpage
\bibliography{references}

\end{document}